\documentclass[superscriptaddress,aps,prb,noeprint,reprint,floatfix]{revtex4-2}
\usepackage{graphicx}
\usepackage{braket}
\usepackage{amssymb}
\usepackage{amsmath}
\usepackage{dcolumn}
\usepackage[utf8]{inputenc}
\usepackage{bm}
\usepackage{hyperref}
\usepackage{stfloats}
\usepackage{color}
\usepackage{xcolor}
\usepackage{float}

\newcommand{\cn}[1]{\raisebox{.5pt}{\textcircled{\raisebox{-.9pt} {#1}}}}

\begin{document}

 % Exibe título, autor e data
\title{Competing spin-1 and spin-2 regimes in a frustrated four-leg spin-1/2 ladder}
\author{D. S. Almeida and R. R. Montenegro-Filho}
\affiliation{Laborat\'{o}rio de F\'{i}sica Te\'{o}rica e Computacional, Departamento de F\'{i}sica, Universidade Federal de Pernambuco, 50760-901 Recife-PE, Brasil}
\date{\today}

\begin{abstract}
We investigate a frustrated four-leg spin-$1/2$ ladder using density matrix
renormalization group calculations. The uniform system displays three regimes: short-range
antiferromagnetic legs, short-range ferromagnetic legs, and an effective spin-2
Heisenberg chain, separated by a crossover and a first-order transition. The spin-2
regime is confirmed through its finite string order parameter, edge-localized
excitations, and excellent agreement with a projected $S_r=2$ effective Hamiltonian.
Recasting the model as two frustrated two-leg ladders coupled by rung and diagonal
interactions, we track how the trivial and Haldane phases of an isolated ladder evolve
as interladder couplings are introduced. The resulting phase diagrams reveal crossover
and first-order lines whose locations are captured by the spin-2 projection and show
how singlet- and triplet-dominated regimes reorganize when two ladders merge into a
four-leg structure, clarifying the emergence of effective spin-1 versus spin-2 behavior.
\end{abstract}

\maketitle

\section{Introduction}

Interacting spin models in low-dimensional systems are particularly sensitive to quantum fluctuations. For instance, spin-$s$ chains are gapped for integer values of $s$ and gapless for half-integer values, as originally proposed by Haldane~\cite{Haldane1983}. Moreover, the (gapped) spin-$1$ chain exhibits a hidden nonlocal order~\cite{Kennedy1992a} and hosts spin-$1/2$ edge states under open boundary conditions. Owing to their distinct entanglement features, the ground states of {odd-integer} spin-$s$ chains realize symmetry-protected topological (SPT) phases~\cite{Wen2019}, whose robust edge excitations and topological signatures have motivated extensive theoretical and experimental investigations in recent years.

Ladder systems~\cite{DagottoScience,evertz1997,l.silva2021,ramos2026}, composed of coupled spin-$s$ chains, provide an ideal framework for exploring the interplay between quantum fluctuations, topology, and dimensional crossover \cite{li2025,wuttig2025,furuya2016}. They interpolate naturally between one-dimensional chains and higher-dimensional quantum magnets, allowing controlled studies of how dimensionality affects magnetic order and excitation spectra~\cite{furuya2016}. In particular, they offer a versatile platform for testing theoretical predictions in both material realizations and synthetic quantum simulators. The ground state of the spin-$1/2$ two-leg ladder, for instance, is well described by a short-range valence bond state, and its topological features can be experimentally engineered~\cite{sompet2022} in optical lattice setups with ultracold atoms. A fundamental property of unfrustrated $n$-leg spin-$1/2$ ladders is the presence of a spin gap when $n$ is even and its absence when $n$ is odd~\cite{MReigrotzki1994,white1994b,sierra1996,PhysRevB.89.094424}. Increasing the number of legs therefore provides a natural route to study the evolution from one-dimensional to two-dimensional quantum magnetism.

From the experimental side, the low-temperature magnetic behavior of various quasi-one-dimensional compounds has been attributed to the presence of ladder-like arrangements in their crystal structures. Notable examples include the copper oxide series Sr$_{x-1}$Cu$_{x+1}$O$_{2x}$ ($x = 3, 5, 7, \ldots$)~\cite{gopalan1994a}, composed of spin-$1/2$ ladders with $n = (x+1)/2$ legs; and the organic compound C$_{28}$H$_{42}$N$_4$O$_4$ (BIP-TENO)~\cite{nomura2022,kohshiro2021,sakai2003a,katoh2002,katoh2000}, which is effectively described by a frustrated spin-$1$ two-leg ladder. In the latter, the spin-$1$ degrees of freedom emerge from two ferromagnetically coupled spin-$1/2$ NO groups, and the magnetization curve displays nontrivial fractional plateaus. Other compounds, such as the weakly coupled two-leg ladders Ba$_2$CuTeO$_6$~\cite{PhysRevB.95.104428,PhysRevB.98.174410,Pughe2023} and C$_9$H$_{18}$N$_2$CuBr$_4$~\cite{PhysRevB.89.174432,PhysRevLett.122.127201,Hong2017}, lie close to a quantum critical point~\cite{PhysRevB.65.092406} separating magnetically ordered and spin-gapped phases. Furthermore, the metal–organic coordination compound (C$_5$H$_9$NH$_3$)$_2$CuBr$_4$~\cite{PhysRevB.110.094101} exhibits evidence of a phase in which its magnetic properties can be effectively described by a two-ladder structure.

Frustration adds a new layer of complexity to these systems. In spin-$1/2$ two-leg frustrated ladder Hamiltonians~\cite{thereza2025,almeida2023b,wessel2017a,Michaud2010,hikihara2010a,fouet2006a,honecker2000a,wang2000a,
weihong1998,Mila1998,Gelfand1991}, frustration leads to competing ground states where the system may behave effectively as a spin-$1$ chain (Haldane phase), and where fractional magnetization plateaus, magnetization jumps, and multicritical points emerge. In particular, frustrated two-leg spin-$s$ ladders~\cite{Michaud2010} can be effectively mapped to a spin-$1/2$ XXZ chain in the strong-rung-coupling regime, and to a spin-$2s$ chain~\cite{USchollwock1995,schollwock_s_1996,PhysRevLett.87.047203,PhysRevB.60.14529,doi:10.7566/JPSJ.87.105002} in the weak-rung limit. {More generally, an unfrustrated anisotropic spin-$S$ Heisenberg chain can be represented by $2S$ ferromagnetically coupled spin-$1/2$ chains, as argued by Schulz~\cite{Schulz1986} through a continuum, weak-coupling analysis of the spin-$S$ chain.} The coupling between two spin-$1/2$ frustrated ladders has also been explored through the coupled cluster method~\cite{JIANG2020123131}, revealing a rich variety of quantum phases and transitions.

Here, we study a 4-leg spin-$1/2$ ladder model with diagonal frustrating couplings inside each plaquette, focusing on two coupling regimes using the density matrix renormalization group (DMRG)~\cite{schollwock2011,Schollwock2005,schollwock2007,white1992,white1993a}. We first describe the phases of the 4-leg ladder in the \textit{uniform regime}, where the rung, leg, and diagonal couplings are constant throughout the system. Next, we vary the couplings between the two 2-leg ladders that compose the 4-leg structure, referred to as the \textit{coupled 2-leg spin-1/2 frustrated ladder}, in order to examine how the phases of the spin-$1/2$ 2-leg frustrated ladder evolve as the two ladders are coupled to form the 4-leg system with uniform interactions.

The interplay between interladder couplings and frustration in quasi-one-dimensional magnets provides an additional motivation
for this class of models. In particular, NMR experiments on the metal–organic compound (C$_7$H$_{10}$N)$_2$CuBr$_4$ (DIMPY)~\cite{jeong2017}
have been interpreted in terms of a hierarchy of effective dimensionalities set by the anisotropic and partially
frustrated interladder coupling network~\cite{furuya2016}. Together with the ladder compounds Ba$_2$CuTeO$_6$ and C$_9$H$_{18}$N$_2$CuBr$_4$
mentioned above, these results indicate that the coupling between frustrated ladders is a relevant ingredient in the
low-energy physics of such systems, motivating the microscopic study of how the ground-state phases of an isolated frustrated
two-leg ladder evolve once an additional interladder coupling is introduced.

This article is organized as follows. The Section~\ref{sec:models} describes the models, the DMRG implementation, and the observables used to characterize the phases. The phase diagram of the uniform 4-leg ladder is discussed in Sec.~\ref{sec:4leg}, followed by an analysis in Sec.~\ref{sec:2legcoupled} of how the 2-leg ladder phases evolve as the interladder couplings are turned on. Finally, Sec.~\ref{sec:summary} summarizes our main results.

\section{Models and Method}
\label{sec:models}

\begin{figure}
    \centering
    \includegraphics[width=0.4\textwidth]{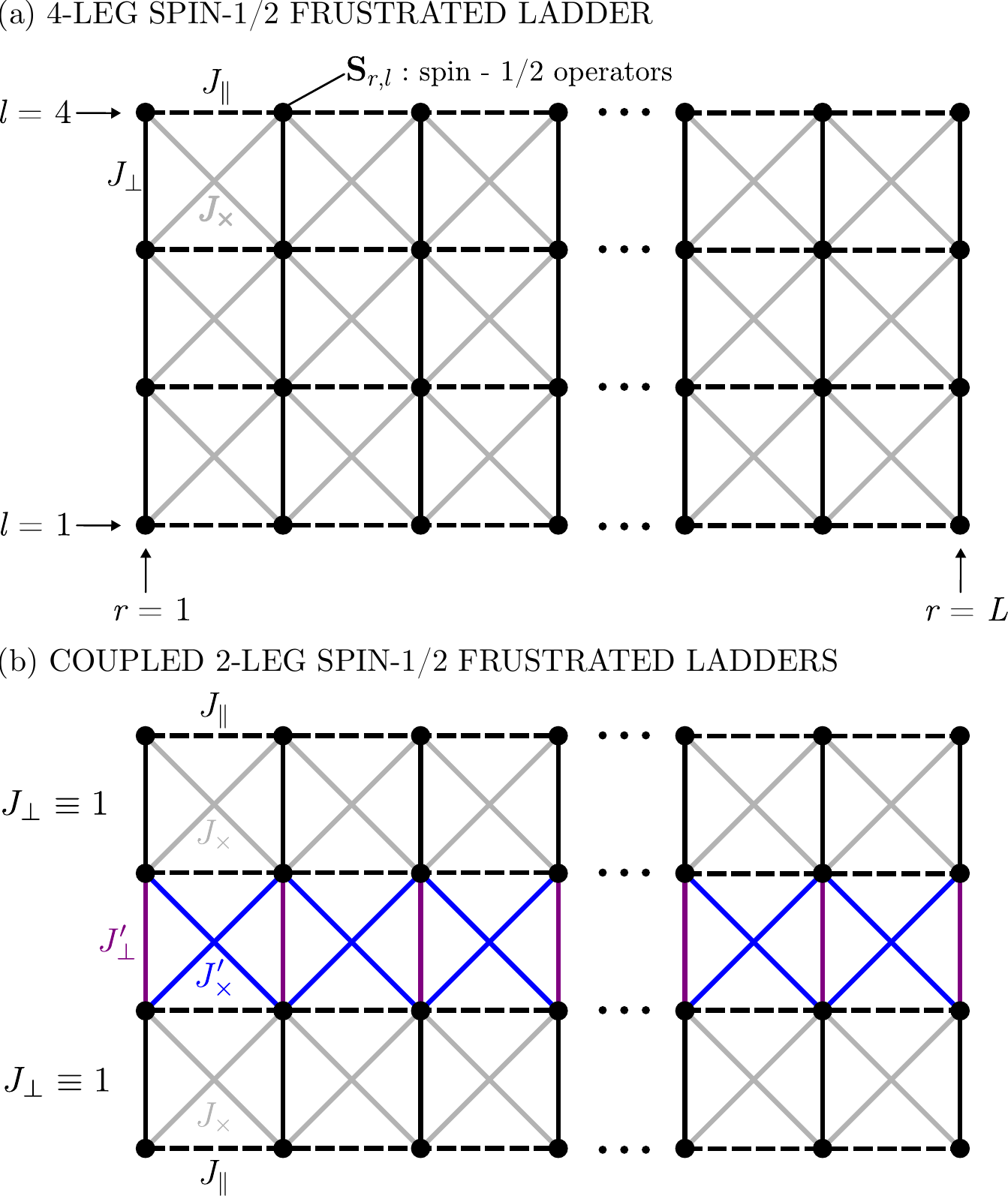} % substitua pelo nome do seu arquivo
    \caption{(a) Schematic representation of the 4-leg spin-$\frac{1}{2}$ frustrated ladder of size $L$. The intra-rung coupling $J_\perp \equiv 1$, the inter-rung couplings are $J_\times$ and $J_\parallel$. The triple rungs are labelled by
    $r$ ($=1\ldots L$) and the legs by $l$ ($=1\ldots4$). (b) Illustration of the Hamiltonian for two coupled spin-$\frac{1}{2}$ frustrated 2-leg ladders. The intra-ladder couplings are $J_\perp (\equiv 1)$, $J_\parallel$, and $J_\times$; while $J^\prime_\perp$ and $J^\prime_\times$ represent the interaction parameters between the ladders.}
    \label{fig:model}
\end{figure}

We investigate two related spin ladders, illustrated in Figs. \ref{fig:model}(a) and (b), both described by the Hamiltonian
\begin{align}
    \mathcal{H}= J_\parallel &\sum_{l=1}^{4}\sum_{r=1}^{L-1}\mathbf{S}_{r,l}\cdot \mathbf{S}_{r+1,l}\nonumber \\
    + &\sum_{l=1}^3\left[J_{\perp,l}\sum_{r=1}^{L}\mathbf{S}_{r,l}\cdot \mathbf{S}_{r,l+1}\right. \nonumber\\
     &\phantom{\sum_{i}} \left. + J_{\times,l}\sum_{r=1}^{L-1}\left(\mathbf{S}_{r,l}\cdot \mathbf{S}_{r+1,l+1}+\mathbf{S}_{r,l+1}\cdot \mathbf{S}_{r+1,l}\right)\right],
\label{eq:ham}
\end{align}
where $\mathbf{S}_{r,l}$ are spin-1/2 operators at the {four-spin} rung $r$ and leg $l$.
The first model, shown in Fig. \ref{fig:model}(a), is a 4-leg spin-1/2 frustrated ladder in which both the intra-rung couplings $J_{\perp,l}$ and the diagonal couplings $J_{\times,l}$ are leg-independent: $J_{\perp,l}=J_\perp\equiv 1$ and $J_{\times,l}=J_\times$.
The second model, Fig. \ref{fig:model}(b), corresponds to two frustrated 2-leg ladders coupled to each other. The internal rungs of each 2-leg ladder have $J_\perp\equiv 1$ and diagonal couplings $J_\times$, while the inter-ladder couplings are $J_\perp^\prime$ (intra-rung) and $J_\times^\prime$ (diagonal). In both models, $J_\parallel$ denotes the coupling along the legs.

\subsection{DMRG and computed observables}

We employ the density matrix renormalization group (DMRG), implemented via the ITensor library~\cite{10.21468/SciPostPhysCodeb.4}, to compute ground-state properties. All simulations were performed with open boundary conditions, using a maximum bond dimension of up to 3000. To balance accuracy and computational cost, we adopted a truncation error cutoff in the range $10^{-7}$–$10^{-6}$. For thermodynamic extrapolations, a linear fit {as function of $1/L$} was performed using data from three system sizes, $L = 48$, $72$, and $96$ rungs.

To help characterize the different phases of the system, we compute the average effective total rung spin, $\bar{S}$. The total spin operator on rung $r$
\begin{align}
    \mathbf{S}_r = \sum_{l=1}^4 \mathbf{S}_{r,l}
\end{align}
{ leads to the squared total spin operator}
\begin{equation}
 \mathbf{S}_r^2 = \sum_{l=1}^4 \mathbf{S}_{r,l}^2 + 2\sum_{1\leq\alpha<\beta\leq4}\mathbf{S}_{r,\alpha}\cdot\mathbf{S}_{r,\beta},
\end{equation}
{ with an average value}
\begin{equation}
 \braket{\mathbf{S}_r^2}=3 + 2\sum_{1\leq\alpha<\beta\leq4}\langle \mathbf{S}_{r,\alpha}\cdot\mathbf{S}_{r,\beta}\rangle,\label{eq_srbar}
\end{equation}
{that can be used to define an effective rung-spin magnitude $\bar{S}_r$ through}
\begin{equation}
 \bar{S}_r(\bar{S}_r+1)\equiv \braket{\mathbf{S}_r^2}.
\end{equation}

The average effective total rung spin over the system is defined as
\begin{equation}
    \bar{S} = \frac{1}{L}\sum_{r=1}^{L}\bar{S}_r.
\label{eq:Smedio}
\end{equation}

To investigate the  presence of hidden order, we compute the string order {correlations, which
for the 4-leg ladder \cite{doi:10.1143/JPSJ.65.560,ueda2008}, is calculated through}

\begin{align}
    \mathcal{O}^z(i,j)
    = \left\langle
    S^z_i \exp\!\left[\frac{i\pi}{\bar{S}}\sum_{k=i+1}^{j-1}S^z_k\right] S^z_j
    \right\rangle,
\end{align}
{ where the rung spin operator is}
\begin{equation}
    S^z_r = \sum_{l=1}^4 S^z_{r,l}.
\label{eq:rungm}
\end{equation}

If the system exhibits a hidden order, the parameter $\mathcal{O}_{\text{str}}^z=\lim_{|i-j|\rightarrow \infty}\mathcal{O}^z(|i-j|)\neq 0$.
For a finite system of size $L$ with open boundaries \cite{doi:10.1143/JPSJ.65.560,ueda2008}, we consider the $L/3$ rungs at the center, such that $i=\frac{L}{3}+1$ and $j=\frac{2L}{3}$, defining the parameter
\begin{equation}
{o}^z(L)\equiv \mathcal{O}^z\left(\frac{L}{3}+1,\frac{2L}{3}\right),
\label{eq:fsstring}
\end{equation}
{ and estimate the hidden order parameter as}
\begin{equation}
\mathcal{O}^z_{\text{str}}=\lim_{L\rightarrow\infty}{o}^z(L).
\label{eq:string}
\end{equation}

This definition reduces to the standard string order parameter of the spin-1 chain when $\bar{S}=1$, and generalizes it to effective rung spins $\bar{S}>1$.
Analogous expressions for $\bar{S}$ and $\mathcal{O}^z_{\text{str}}$ can be defined for the 2-leg ladder case.

Finally, we analyze the spin–spin correlation function along a leg, defined as
\begin{align}
    C_l(d) = \big\langle \!\big\langle \mathbf{S}_{r,l}\cdot \mathbf{S}_{s,l}\big\rangle\!\big\rangle_{|r-s|=d},
\label{eq:corr}
\end{align}
where the inner brackets $\langle \cdots \rangle$ denote the quantum expectation value, and the outer brackets $\langle \cdots \rangle_{|r-s|=d}$ represent an average over all site pairs $(r,l)$ and $(s,l)$ separated by a distance $d=|r-s|$.
This averaging procedure reduces boundary effects and provides a smoother measure of correlations along each leg.

\begin{figure}
    \centering
    \includegraphics[width=0.4\textwidth]{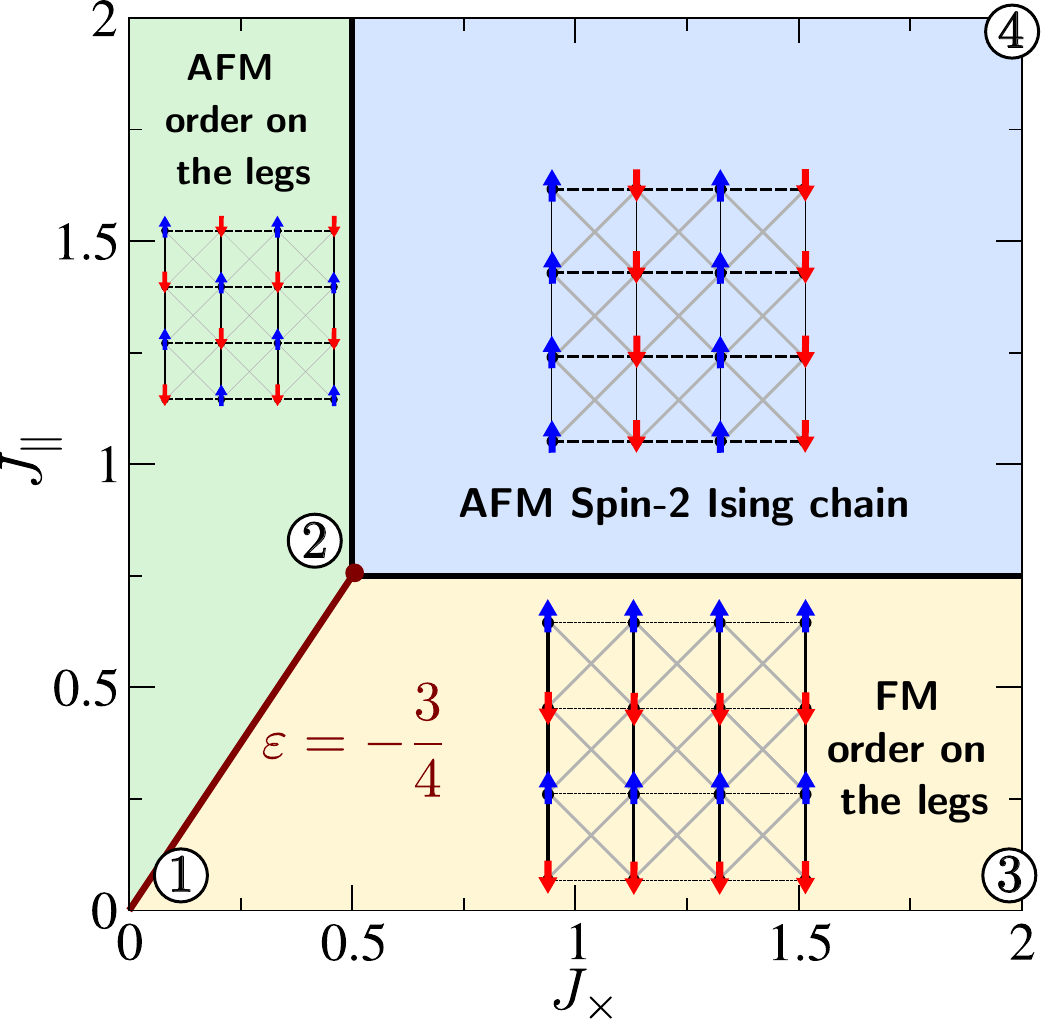} % substitua pelo nome do seu arquivo
    \caption{Phase diagram of the Ising limit. For $J_\parallel=\tfrac{3}{2}J_\times$, the $J_\times$ couplings exactly cancel the $J_\parallel$ ones, so the energy per rung is $\varepsilon=-\tfrac{3}{4}$ and depends only on $J_\perp\equiv1$. In the green region, there is antiferromagnetic (AFM) order on the legs; in the blue region, each rung has a total spin-$2$, with AFM order on the legs; in the yellow region, the ground state shows ferromagnetic (FM) order on the legs. Markers \cn{1}--\cn{4} denote the parameter sets $(J_\times,J_\parallel)$: \cn{1} $(0.1,0.1)$, \cn{2} $(0.4,0.85)$, \cn{3} $(2,0.1)$, \cn{4} $(2,2)$.}
    \label{fig:classic}
\end{figure}

\section{4-leg spin-1/2 frustrated Ladder}
\label{sec:4leg}

Here we discuss our results for the model illustrated in Fig.~\ref{fig:model}(a), with constant values of $J_\parallel$, $J_\perp \equiv 1$, and $J_\times$ along the lattice.

\subsection{Ising limit}

{The ordered phases of the Ising limit are easily obtainable configurations, which we use only as a convenient reference to organize the discussion of the corresponding Heisenberg model.}
{In this limit, we consider only the $z$ components of the Hamiltonian in Eq.(\ref{eq:ham}):
\begin{align}
    {H}= J_\parallel &\sum_{l=1}^{4}\sum_{r=1}^{L-1}{S^z}_{r,l}{S^z}_{r+1,l}\nonumber \\
    + &\sum_{l=1}^3\left[J_{\perp,l}\sum_{r=1}^{L}{S^z}_{r,l}{S^z}_{r,l+1}\right. \nonumber\\
     &\phantom{\sum_{i}} \left. + J_{\times,l}\sum_{r=1}^{L-1}\left({S^z}_{r,l}{S^z}_{r+1,l+1}+{S^z}_{r,l+1}{S^z}_{r+1,l}\right)\right].
\label{eq:hamising}
\end{align}
}
In the ground state of the Ising limit, three ordered phases emerge, as sketched in Fig.~\ref{fig:classic}, with energies per rung ($\varepsilon$) given by
\begin{align}
 \varepsilon_{\parallel} &= -\frac{3}{4} - J_\parallel + \frac{3}{2}J_\times,\\
 \varepsilon_{\times} &= -\frac{3}{4} + J_\parallel - \frac{3}{2}J_\times,\\
 \varepsilon_{\text{spin-2}} &= \frac{3}{4} - J_\parallel - \frac{3}{2}J_\times.
\end{align}
The first phase, with energy $\varepsilon_{\parallel}$, exhibits antiferromagnetic (AFM) correlations along the legs and is stabilized when $J_\perp$ and $J_\parallel$ are the dominant couplings (green region in Fig.~\ref{fig:classic}). In contrast, for $J_\perp$ and $J_\times$ sufficiently larger than $J_\parallel$ (yellow region), the system displays ferromagnetic (FM) order along the legs, with energy $\varepsilon_\times$. In both of these phases, the total magnetization on each rung is zero.

Along the line $J_\parallel = \tfrac{3}{2}J_\times$, the energies of these two phases become degenerate, $\varepsilon = -\tfrac{3}{4}$, defining the phase boundary between the AFM and FM leg-ordered states. Finally, when $J_\parallel$ and $J_\times$ are sufficiently stronger than $J_\perp$ (blue region), the spins within each rung align ferromagnetically, and the system behaves as an effective antiferromagnetic spin-2 chain, with energy $\varepsilon_{\text{spin-2}}$.

In the parameter space $(J_\times, J_\parallel)$ of Fig.~\ref{fig:classic}, we have indicated several representative points: \cn{1} $(0.1, 0.1)$, \cn{2} $(0.4, 0.85)$, \cn{3} $(2, 0.1)$, and \cn{4} $(2, 2)$, which will be discussed below for the Heisenberg model. In the Heisenberg case, the presence of quantum fluctuations leads to short-range correlations rather than perfect magnetic alignment. Furthermore, the phase associated with $\varepsilon_{\text{spin-2}}$ becomes particularly interesting: the ferromagnetic alignment within each rung forms effective spin-2 units, which couple antiferromagnetically along the legs, giving rise to a Haldane-type phase characterized by a nonvanishing string order parameter and a finite spin gap. {The reduction of the easy-axis anisotropy is known to connect the Ising-N\'eel state to the Haldane phase in the single spin-1 chain~\cite{BotetJullienKolb1983}. In the present model, however, the spin-2 unit is not rigid: the rung spin $S_r$ is emergent, so whether the Ising spin-2 region survives as a single spin-2 Heisenberg chain or gives way to regimes with other $S_r$ is set by the competition among the couplings, as analyzed below.} The evolution of these phases and the corresponding quantum phase boundaries are analyzed in detail below.

\subsection{Heisenberg model}

\begin{figure}
    \centering
    \includegraphics[width=0.4\textwidth]{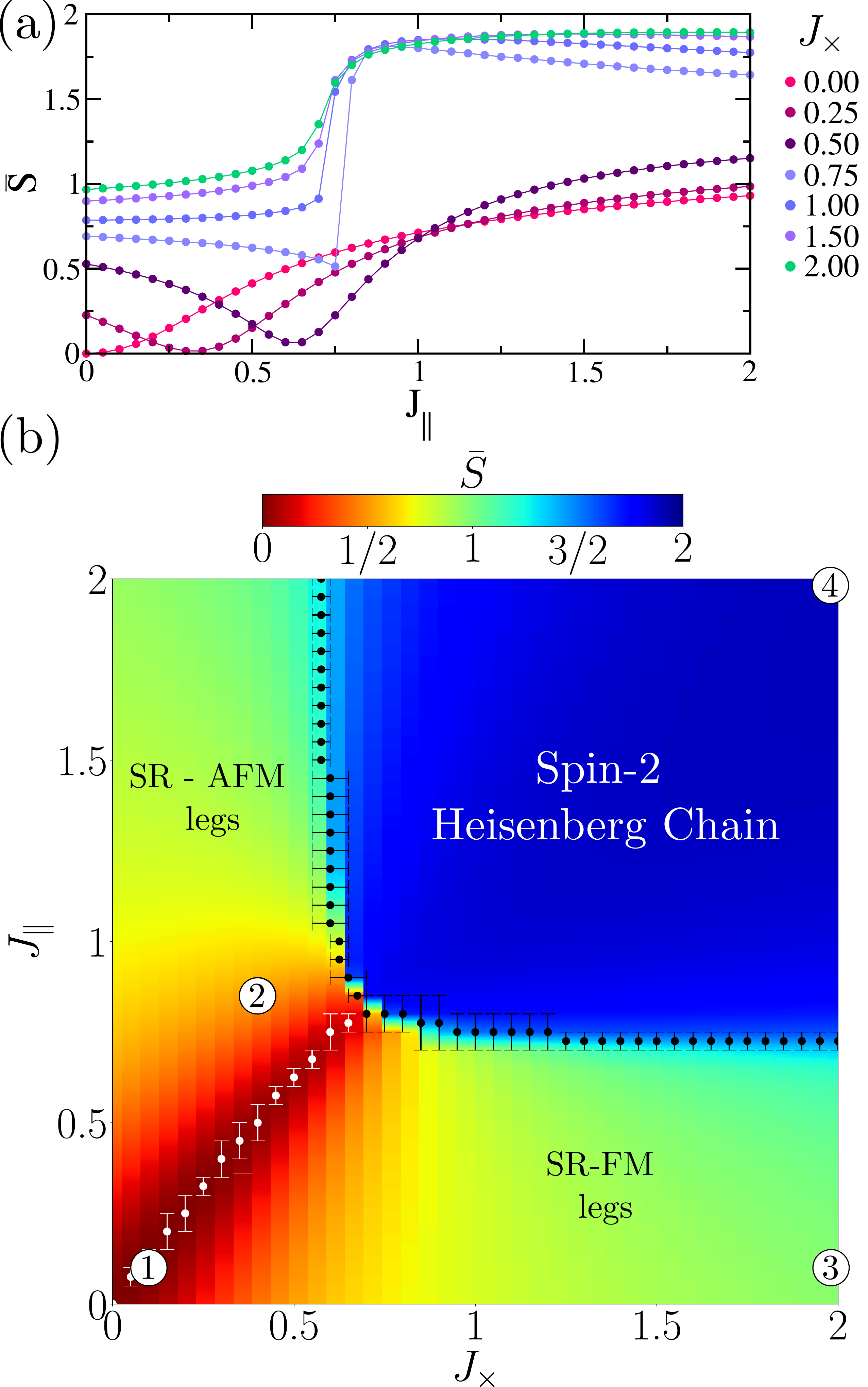} % substitua pelo nome do seu arquivo
    \caption{Ground-state phase diagram of the frustrated spin-1/2 four-leg ladder in the $(J_\times, J_\parallel)$ plane, obtained from DMRG calculations for a system with $L = 32$ rungs. (a) Average total rung spin $\bar{S}$ plotted along vertical cuts of the $(J_\times, J_\parallel)$ plane. For $J_\times \leq 0.5$, the curves exhibit a minimum marking a crossover, while for $J_\times \geq 0.75$ a jump occurs, indicating a first-order transition. (b) Phase diagram, where the color map represents $\bar{S}$ evaluated over the central 24 rungs to minimize boundary effects. White circles mark the crossover line between the phase with short-range (SR) antiferromagnetic (AFM) correlations and that with SR ferromagnetic (FM) correlations along the legs. Along this line, the dominant correlations occur between spins on the same rung. Black circles denote the first-order transition separating the effective spin-2 Heisenberg chain from the SR-FM and SR-AFM phases on the legs. We indicate the points \cn{1} $(0.1, 0.1)$, \cn{2} $(0.4, 0.85)$, \cn{3} $(2, 0.1)$, and \cn{4} $(2, 2)$ for further reference.
    }
    \label{fig:diagram}
\end{figure}

In Fig.~\ref{fig:diagram}(a), we present the average total rung spin $\bar{S}$, defined in Eq.~(\ref{eq:Smedio}), as a function of $J_\parallel$ for the indicated values of $J_\times$. Several features are noteworthy. For $J_\times \leq 0.5$, $\bar{S}$ exhibits a minimum with $\bar{S}\approx 0$, corresponding to nearly decoupled rungs in local singlet states. In contrast, for $J_\times \geq 0.75$, a jump occurs, signaling a first-order transition to a state with $\bar{S}\approx 2$. The magnitude of this discontinuity decreases gradually as $J_\times$ increases. In Fig.~\ref{fig:diagram}(b), $\bar{S}$ is displayed as a color map over a grid of points in the $(J_\times, J_\parallel)$ parameter space. The white symbols mark the points where the minimum in $\bar{S}$ is observed, an almost-isolated-rung regime, delineating a crossover line between phases exhibiting short-range antiferromagnetic (SR-AFM) and short-range ferromagnetic (SR-FM) correlations along the legs. The black symbols indicate the first-order transition line separating the effective spin-2 Heisenberg chain from the SR-FM and SR-AFM legs regimes. {An analytical estimate for the transition from the
almost-isolated-rung regime ($\bar{S}\approx 0$) to the effective
spin-2 Heisenberg chain, based on a comparison of rung ground-state
energies, in the spirit of Gelfand's framework~\cite{Gelfand1991},
is presented in Appendix~\ref{app:transition}}.
Below, we characterize in detail the SR-AFM and SR-FM legs phases, as well as the spin-2 Heisenberg chain regime.

\begin{figure}
    \centering
    \includegraphics[width=0.4\textwidth]{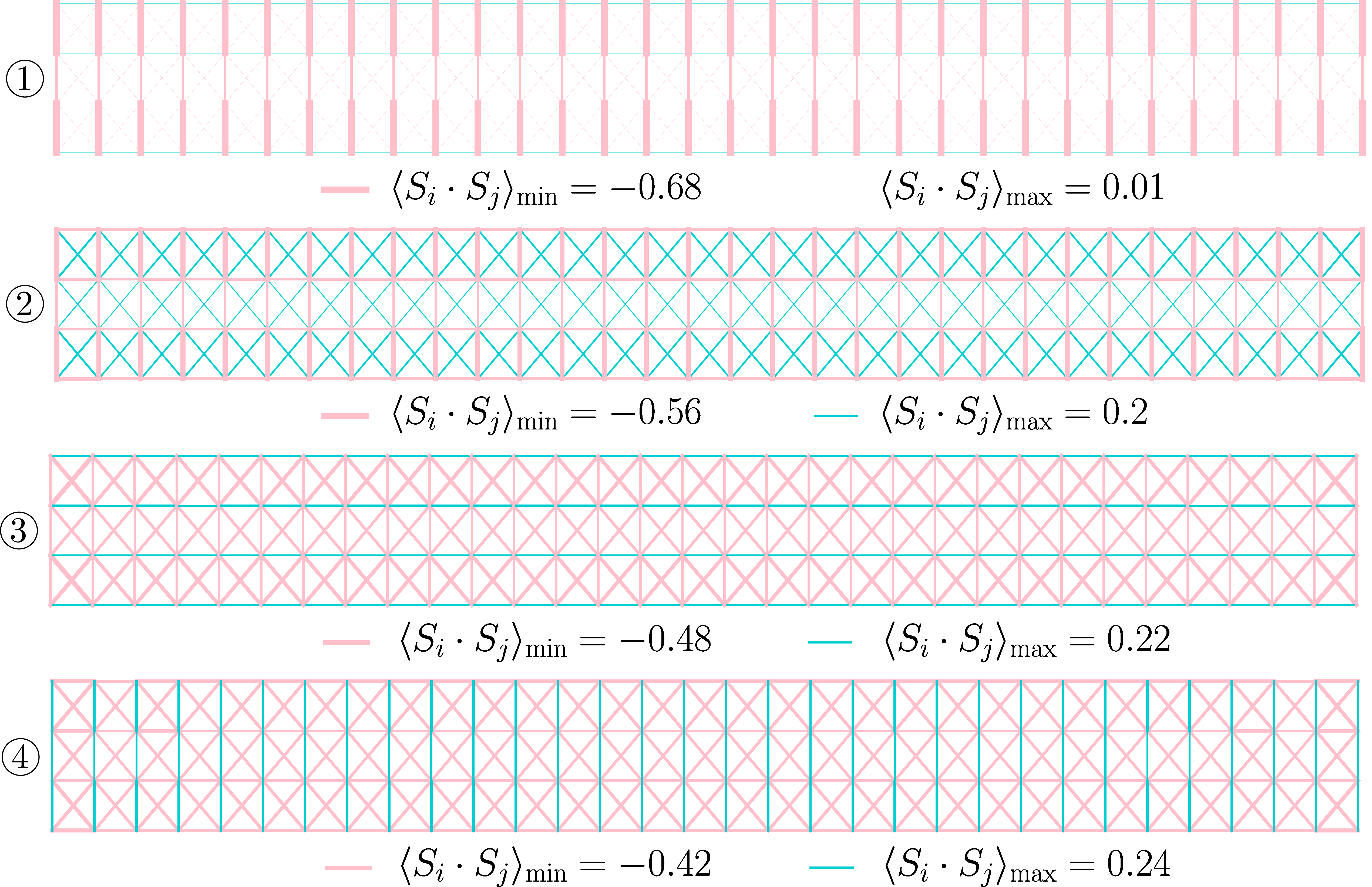}
    \caption{Spin–spin correlations $\langle \mathbf{S}_i\cdot \mathbf{S}_j\rangle$ between spins connected by $J_\perp$, $J_\times$, and $J_\parallel$, for the points labeled in Fig. \ref{fig:diagram}(b) and a system with 32 rungs. The line widths are proportional to the magnitude of $\langle \mathbf{S}_i\cdot \mathbf{S}_j\rangle$, using as references the values indicated below each configuration.}
    \label{fig:diagramcorr}
\end{figure}

Representative spin–spin correlations for the four labeled points in Fig.~\ref{fig:diagram}(b) are shown in Fig.~\ref{fig:diagramcorr} for the central sites of a ladder with 32 rungs. For \cn{1} $(0.1, 0.1)$, corresponding to the SR-FM leg phase, the correlations are strongest between spins on the same rung, with ferromagnetic correlations along the legs and antiferromagnetic correlations along the diagonals.
For \cn{2} $(0.4, 0.85)$, in the SR-AFM legs phase, the rung correlations weaken, while nearest-neighbor correlations along the legs become antiferromagnetic and those along the diagonals become ferromagnetic.
Point \cn{3} $(2, 0.1)$, also belonging to the SR-FM legs phase, displays correlations similar to \cn{1}, but with a stronger ferromagnetic component ($\approx 0.22$) along the legs and more uniform correlations within each rung.
Finally, for \cn{4} $(2, 2)$, associated with the spin-2 phase, the spins on each rung exhibit triplet-like correlations ($\approx 0.25$), consistent with $\bar{S}\approx 2$, and both leg and diagonal couplings are satisfied within the effective spin-2 chain structure.

\begin{figure}
    \centering
    \includegraphics[width=0.45\textwidth]{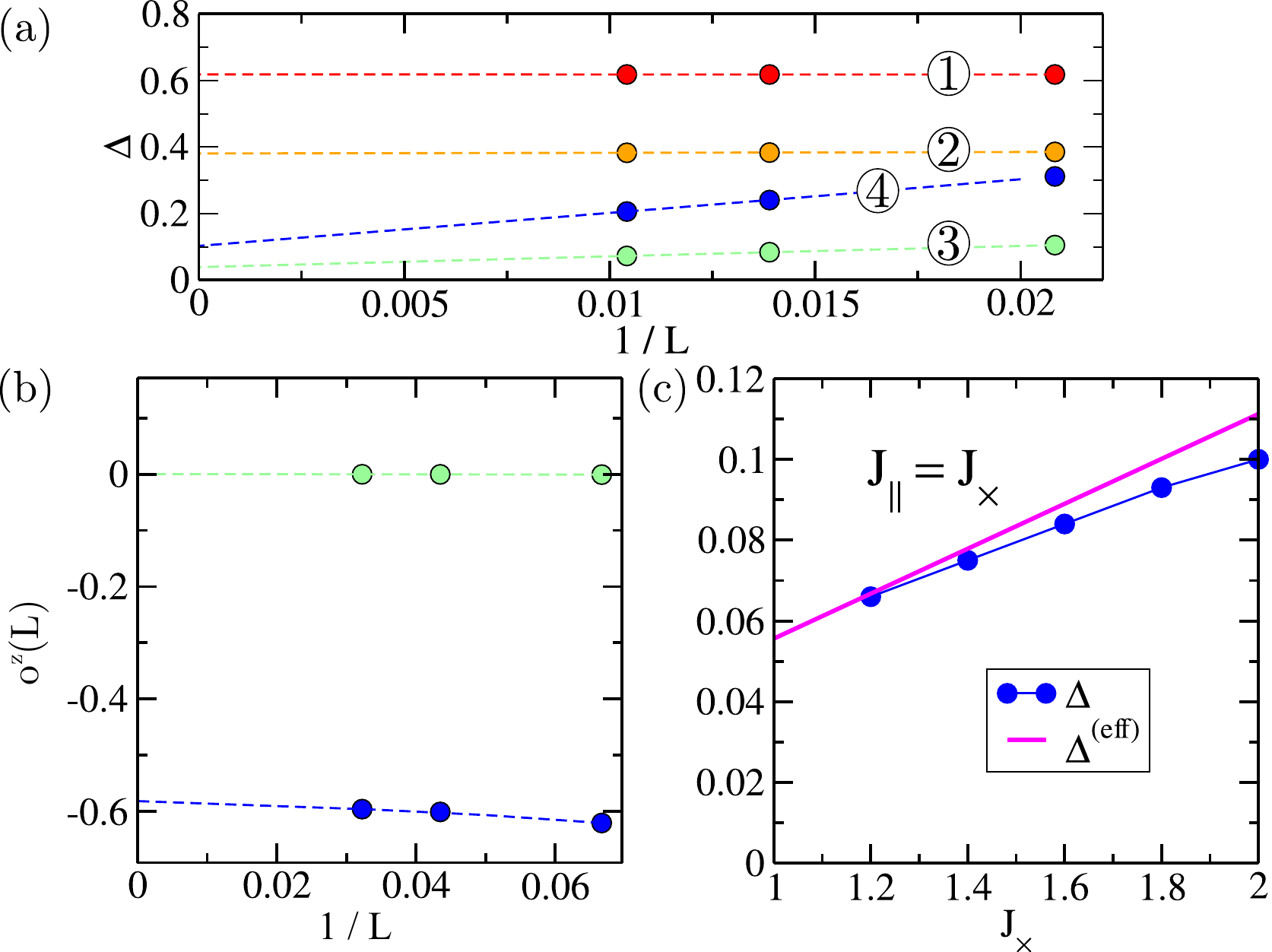} % substitua pelo nome do seu arquivo
    \caption{ (a) Linear extrapolation of the spin gap $\Delta$ for the $(J_\times, J_\parallel)$ values indicated in Fig.~\ref{fig:diagram}(b).  {(b) Linear extrapolation of the finite-size string order parameter $o^z$, Eq. (\ref{eq:fsstring}), for points \cn{3} and \cn{4}}. (c) Comparison between the extrapolated ladder gap $\Delta$ obtained from DMRG and the effective spin-2 prediction $\Delta^{(\mathrm{eff})} = 0.089 \times \tfrac{5}{8}J_\times$ along the line $J_\parallel = J_\times$. DMRG data are extrapolated using system sizes $L = 48, 72, 96$.}
    \label{fig:gapstring}
\end{figure}

\begin{figure}
    \centering
    \includegraphics[width=0.4\textwidth]{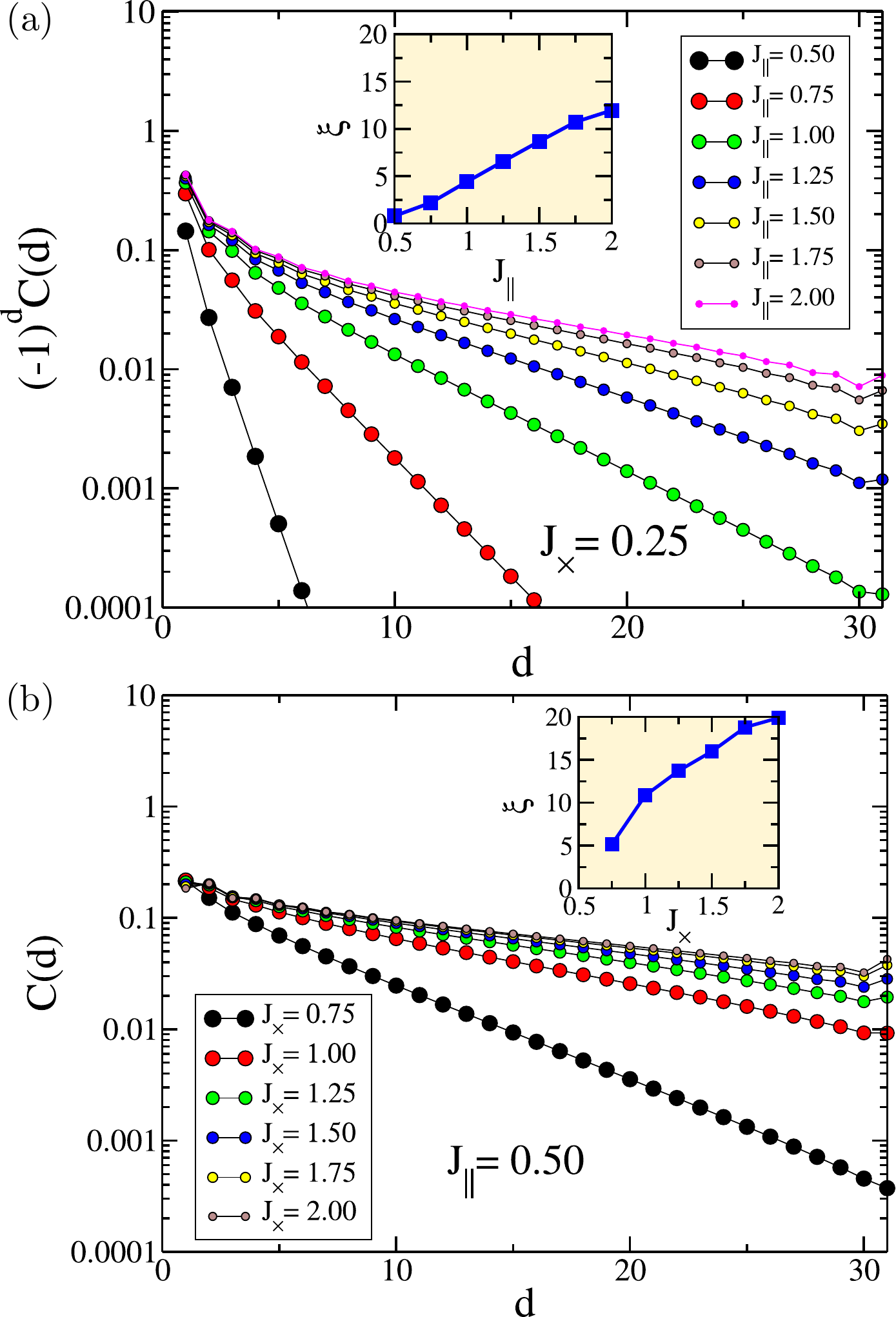} % substitua pelo nome do seu arquivo
    \caption{Spin-spin correlations along the $l=1$ leg, $C(d)$, as a function of the distance $d$ between the spins {for systems of size $L=32$}. (a) In the phase SR-AFM legs phase: $(-1)^d C(d)$ for $J_\times=0.25$ fixed and the indicated values of $J_\parallel$; (b) In the phase SR-FM legs phase: $C(d)$ for $J_\parallel=0.50$ fixed and the indicated values of $J_\times$. {The oscillatory behavior for the largest values of $d$ is a finite-size effect.} Insets show the correlation length $\xi$ as a function of (a) $J_\parallel$ and (b) $J_\times$ for the data shown in the corresponding main figures.}
    \label{fig:correlations}
\end{figure}

In Fig.~\ref{fig:gapstring}(a), we present the finite-size scaling analysis of the spin gap for the points \cn{1}, \cn{2}, \cn{3}, and \cn{4} indicated in the phase diagram of Fig.~\ref{fig:diagram}(b). The bulk energy gap is defined as the energy difference between the $S^z=0$ and $S^z=1$ sectors for the first three points. For point \cn{4}, however, we use the difference between the $S^z=3$ and $S^z=2$ sectors, since in this regime the system is effectively described by a spin-2 chain with spin-1 edge states. {The extrapolated values of the gap $\Delta$ indicate gapped phases, with $\Delta_{\cn{1}}=0.62$, $\Delta_{\cn{2}}=0.38$, $\Delta_{\cn{3}}=0.04$, and $\Delta_{\cn{4}}=0.10$ in the thermodynamic limit.} {The tiny value of $\Delta_{\cn{4}}$ can be understood by its effective spin-2 chain character, which we develop below. For point $\cn{3}$, in the SR-FM legs phase, the finite value of the gap is reinforced by
the transverse spin correlation function, discussed in Fig. \ref{fig:correlations}, which exhibits an exponential behavior.}

In particular, Fig.~\ref{fig:gapstring}(b) shows that the string order parameter $\mathcal{O}^z_{\text{str}}$, defined in Eq.~(\ref{eq:string}), remains finite  for point \cn{4}, with $\mathcal{O}^z_{\text{str}}\approx -0.6$, a value consistent with the spin-2 chain result \cite{USchollwock1995,schollwock_s_1996}, $\mathcal{O}^z_{\text{str}}=-0.73$. By contrast, $\mathcal{O}^z_{\text{str}}$ vanishes in the other phases, as illustrated for point \cn{3} in Fig.~\ref{fig:gapstring}(b). {For the points \cn{1} and \cn{2}, the behavior of $o^z(L)$ as a function of $1/L$ is similar
to that of the point \cn{3}, such that $\mathcal{O}^z_{\text{str}}=0$ for these points}.

The effective description of the spin-2 phase becomes more transparent by projecting the Hamiltonian onto the subspace of spin-2 states on each rung. The projection of the four spin-1/2 operators on the same triple rung $r$ onto the subspace with total spin $S_r=2$ (or, equivalently, using a semiclassical argument) implies
\begin{equation}
\mathbf{S}_{r,l} = \frac{1}{4}\,\mathbf{T}_r,
\label{eq:projection}
\end{equation}
where $\mathbf{T}_r$ denotes the effective spin-2 operator. Projecting the Hamiltonian onto the subspace where all triple rungs are in the $S_r=2$ sector yields
\begin{equation}
    \mathcal{H}_{\{S_r=2\}} = \text{const.} + J_{\mathrm{eff}}^{(\mathrm{spin\text{-}2})} \sum_{r} \mathbf{T}_r \cdot \mathbf{T}_{r+1},
\label{eq:Hspin2eff}
\end{equation}
with the coupling of the effective spin-2 chain given by
\begin{equation}
J_{\mathrm{eff}}^{(\mathrm{spin\text{-}2})} = \frac{1}{4}J_\parallel + \frac{3}{8}J_\times.
\label{eq:Jeffspin2}
\end{equation}
In particular, the excitation gap of the spin-2 Heisenberg chain is estimated in Ref.~\cite{doi:10.7566/JPSJ.87.105002} as
\begin{equation}
\Delta_{\mathrm{spin\text{-}2}} = 0.089\,J_{\mathrm{eff}}^{(\mathrm{spin\text{-}2})}.
\end{equation}
Neglecting higher-order energy corrections, we therefore expect the bulk excitation gap of the four-leg frustrated ladder in the spin-2 regime to be
\begin{equation}
\Delta^{(\text{eff})} = 0.089\left(\frac{1}{4}J_\parallel + \frac{3}{8}J_\times\right).
\label{eq:gapeff}
\end{equation}

Figure~\ref{fig:gapstring}(c) compares $\Delta^{(\text{eff})}$ along the line $J_\parallel=J_\times$ with the DMRG results for the full model, $\Delta=E(S^z=3)-E(S^z=2)$. The agreement between the two approaches is excellent, with a small deviation emerging at larger $J_\times$, indicating that the spin-2 effective description becomes less accurate as both $J_\times$ and $J_\parallel$ increase. {Thus, in this region the four-leg frustrated ladder realizes spin-2 Heisenberg-chain physics, in line with the qualitative spin-$S$ to $2S$ spin-$1/2$ correspondence discussed by Schulz~\cite{Schulz1986}; the quantitative agreement of the gap further benchmarks the effective spin-2 coupling. We emphasize, however, that this correspondence holds only within the spin-2 region, since across the remainder of the phase diagram the rung multiplet is not locked to $S_r=2$.}

In Fig.~\ref{fig:correlations}, we present the correlation functions $C_{l=1}(d)$ in the SR-AFM and SR-FM legs phases, computed along leg 1 using Eq.~(\ref{eq:corr}). In Fig.~\ref{fig:correlations}(a), corresponding to the SR-AFM legs phase, we fix $J_\times = 0.25$ and vary $J_\parallel$ from $0.5$, close to the almost-isolated-rung line, up to $2.0$.
The exponential decay of $(-1)^d C_{l=1}(d)$ is evident, and the correlation length $\xi$ increases with $J_\parallel$, as shown in the inset of the figure. In Fig.~\ref{fig:correlations}(b), corresponding to the SR-FM legs phase, we fix $J_\parallel = 0.5$ and vary $J_\times$ from $0.75$, near the almost-isolated-rung line, up to $2.0$.
Here, the correlation function also exhibits an exponential form with an increasing correlation length $\xi$ as $J_\times$ increases; however, the saturation value of $\xi$ is larger in this case. The smaller saturation value of $\xi$ in the SR-AFM legs phase can be attributed to the stronger quantum fluctuations characteristic of antiferromagnetically correlated spin systems.

\begin{figure}
    \centering
    \includegraphics[width=0.4\textwidth]{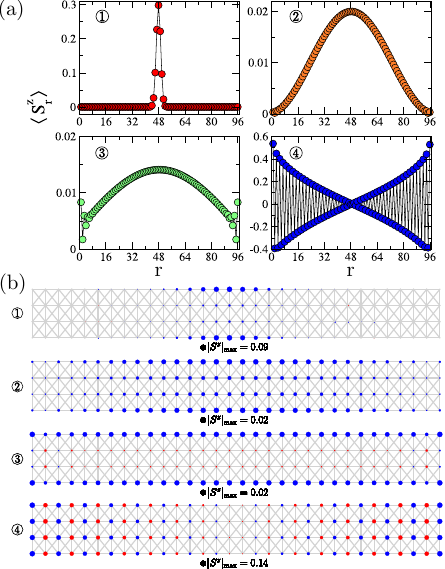}
    \caption{(a) DMRG results for the average rung spin $\langle S^z_r \rangle = \sum_{l=1}^4\langle S^z_{r,l} \rangle$
    for the total spin sector $S^z=1$ along a four-leg ladder with 96 rungs and open boundary conditions.
    (b) DMRG results for the local magnetization profile, $\langle S^z_i \rangle$, on a frustrated spin-$\frac{1}{2}$ four-leg ladder in the total spin sector $S^z_{\text{total}}=1$ {for systems of size $L=32$}. At each site $i$, the circle area is proportional
    to $|\langle S^z_i \rangle|$, considering the value of the reference circle below each panel, and its color
    denotes the sign (blue: $\langle S^z_i \rangle > 0$, red: $\langle S^z_i \rangle < 0$).
    The panels correspond to different parameter points indicated in the phase diagram of Fig.~\ref{fig:diagram}(b).}
    \label{fig:microscopic_ex}
\end{figure}

The excited states in the $S^z = 1$ sector for points \cn{1}, \cn{2}, \cn{3}, and \cn{4} are shown in Fig.~\ref{fig:microscopic_ex}. Figure~\ref{fig:microscopic_ex}(a) displays the rung magnetization $\langle S^z_r \rangle$, defined in Eq.~(\ref{eq:rungm}), as a function of $r$, while Fig.~\ref{fig:microscopic_ex}(b) presents the local magnetization at each site of the ladder. For point~\cn{1}, we observe a highly localized excitation, consistent with the large energy gap and the very short correlation length. In particular, Fig.~\ref{fig:microscopic_ex}(b) shows that the magnetization is predominantly distributed along the outer legs ($l = 1$ and $l = 4$) of the ladder, which minimizes, on average, the number of neighboring spins with the same orientation near this almost-isolated-rung regime.
For point~\cn{2}, the magnon extends over a much larger region of the system, and the magnetization is higher at the central sites compared to point~\cn{1}. For point~\cn{3}, owing to the relatively large correlation length, the bulk excitation spreads throughout the entire ladder, {and presents an oscillatory behavior at the ends of the chain, which we attribute to a boundary effect}. However, as seen in Fig.~\ref{fig:microscopic_ex}(b), it remains strongly localized on the outer legs, for a similar reason as in the \cn{1} case, to minimize the probability of parallel spins connected by $J_\times$.
Consistent with the effective spin-2 phase, which hosts spin-1 edge states, the excitation in the $S^z = 1$ sector for point~\cn{4} is localized near the ladder edges. In this case, we note that the spin-2 chain has a correlation length $\xi \sim 50$~\cite{USchollwock1995,schollwock_s_1996}, consistent with the deep penetration of the edge excitation into the bulk shown in Fig.~\ref{fig:microscopic_ex}(a). Furthermore, the nearly uniform magnetization across sites of the same rung, visible in Fig.~\ref{fig:microscopic_ex}(b), also supports the effective description of the ladder as a spin-2 chain in this regime.

\section{Coupled 2-leg spin-1/2 frustrated Ladders}
\label{sec:2legcoupled}

Here we discuss results for the model illustrated in Fig.~\ref{fig:model}(b), in which
the couplings $J_\perp^\prime$ and $J_\times^\prime$ between the upper and lower
two-leg ladders are varied. Each ladder is characterized by the parameters $J_\times$,
$J_\parallel$, and $J_\perp$ ($\equiv 1$). For an isolated frustrated two-leg ladder, an
SPT spin-1 Haldane phase is realized for certain
combinations of $J_\times$, $J_\parallel$, and $J_\perp$. In particular, for $J_\parallel = 0.8$
and zero magnetic field \cite{almeida2023b}, a topologically trivial paramagnetic state is observed for
$J_\times \lesssim 0.7$, and an SPT spin-1 phase appears for $J_\times \gtrsim 0.7$.
Our goal here is to investigate how these single-ladder states evolve
into the phases of
the four-leg frustrated ladder, Fig.~\ref{fig:diagram}(b), characterized by
$J_\perp^\prime = J_\perp$ and $J_\times^\prime = J_\times$, as the interladder
couplings $J_\perp^\prime$ and $J_\times^\prime$ are gradually turned on.

\begin{figure}
    \centering
    \includegraphics[width=0.4\textwidth]{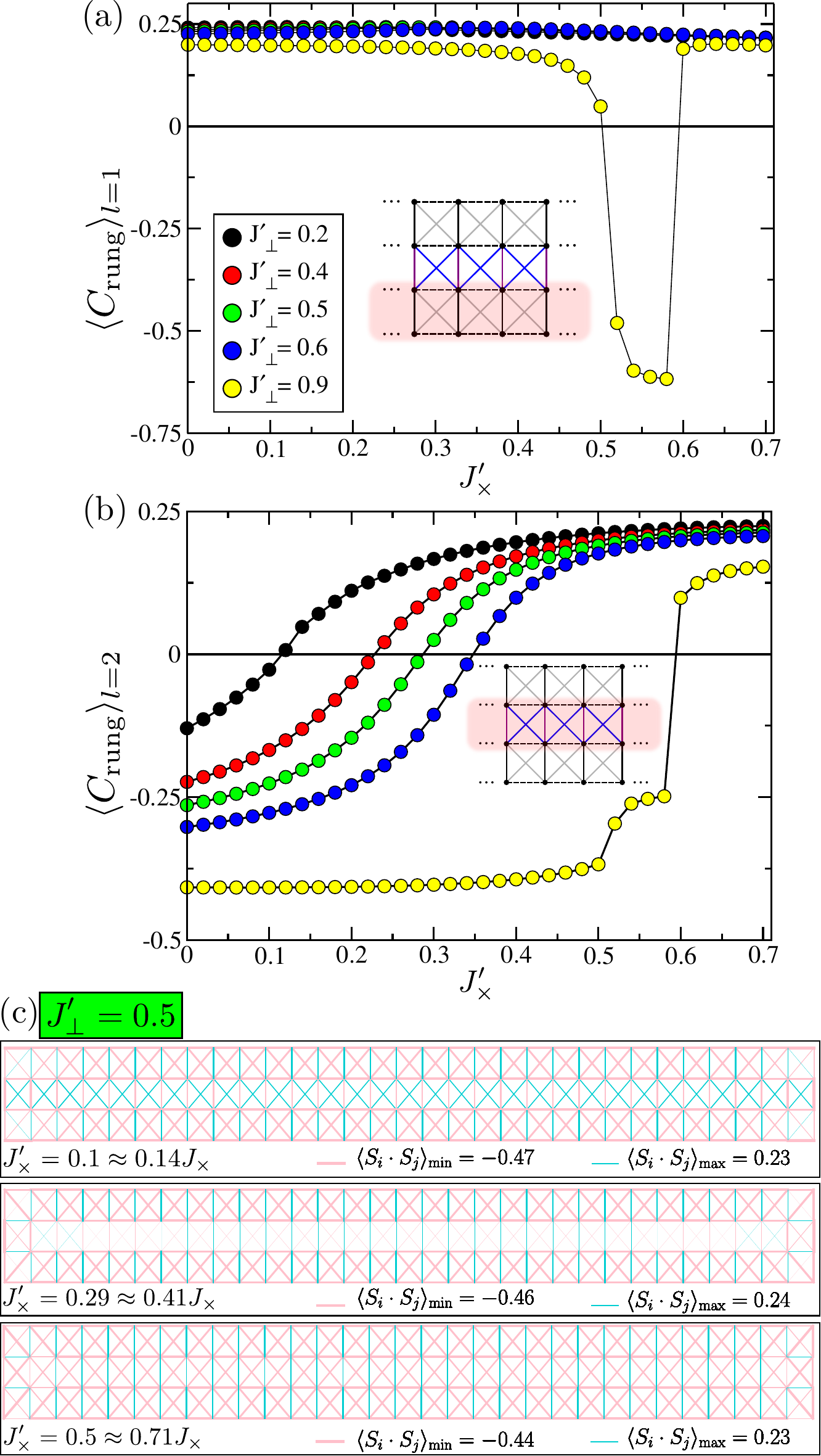}
    \caption{DMRG results for rung spin–spin correlations.
(a) Average rung correlation $\langle C_{\text{rung}} \rangle_{l=1}$ on the lower two-leg ladder as a function of $J_\times'$.
(b) Average rung correlation $\langle C_{\text{rung}} \rangle_{l=2}$ evaluated over the central rungs.
In (a) and (b), the highlighted bonds indicate the rungs included in each average. Blue circles mark the point at which $\langle C_{\text{rung}} \rangle_{l=2}=0$. Data correspond to $L = 32$, $J_\perp=1$, $J_\parallel = 0.8$, and $J_\times = 0.7$. {(c) Spin–spin correlations $\langle \mathbf{S}_i\cdot \mathbf{S}_j\rangle$ between spins connected by $J_\perp$, $J_\times$, and $J_\parallel$, in a system with $L=32$, for $J_\perp'=0.5$
and the indicated values of $J_\times'$. The line widths are proportional to the magnitude of $\langle \mathbf{S}_i\cdot \mathbf{S}_j\rangle$, using as references the values indicated below each configuration.}}
    \label{fig:coupledcorr}
\end{figure}

In Fig.~\ref{fig:coupledcorr}, we present the average rung correlation
\begin{equation}
 \langle C_{\text{rung}}\rangle_l=\frac{1}{L-8}\sum_{r=5}^{L-4}
 \langle \mathbf{S}_{r,l}\cdot \mathbf{S}_{r,l+1}\rangle,
 \label{eq:avrungcorr}
\end{equation}
evaluated over the central rungs ($l=2$) and over the rungs of the lower two-leg ladder ($l=1$), which by symmetry is equivalent to $\langle C_{\text{rung}}\rangle_{l=3}$.
To minimize boundary effects, four rungs at each end of the system are discarded in Eq.~(\ref{eq:avrungcorr}).

In Fig. \ref{fig:coupledcorr}, we fix $J_\parallel=0.8$, $J_\times=0.7$, and $J_\perp=1$, a parameter combination that places an isolated two-leg ladder
($J_\perp'=J_\times'=0$) in the spin-1 SPT phase. We then vary the interladder couplings $J_\times'$ and $J_\perp'$.
For $J_\perp'=J_\perp=1$ and $J_\times'=J_\times=0.7$, the system corresponds to a point of the phase diagram of the four-leg frustrated ladder [Fig.~\ref{fig:diagram}(b)], lying very close to the transition between the spin-2 phase and the SR-FM legs phase.

Figures~\ref{fig:coupledcorr}(a) and (b) show a clear jump at $(J_\perp',J_\times')=(0.9,0.6)$, signaling that the system crosses the first-order transition line of the four-leg ladder.
This jump occurs at a slightly smaller value of $J_\times'$ than $J_\times=0.7$ because $J_\perp'$ is slightly reduced from its 4-leg uniform value $J_\perp=1$. In fact, we can observe in Fig. \ref{fig:coupledcorr}(a)
that there are two jumps as $J_\times$ is reduced, one (stronger) from a correlation $\approx 0.25$ to a value of $\approx -0.63$, and the other (weaker) back to $\approx 0.25$. We notice that the phase between these two jumps exhibit strong antiferromagnetic correlations, similarly to that observed in the state \cn{2} of Fig. \ref{fig:correlations}, due
to its proximity to the crossover line of Fig. \ref{fig:diagram}(b).

For $J_\perp' < 0.9$ in Figs. \ref{fig:coupledcorr}(a) and (b), the curves are smooth for both average rung correlations.
In this regime, weakening $J_\perp'$ while keeping $J_\parallel$ and $J_\times$ relatively strong favors the formation of triplet states on the rungs, as in the case of two nearly independent two-leg ladders.
This behavior is evident in Fig.~\ref{fig:coupledcorr}(a), where $\langle C_{\text{rung}} \rangle_{l=1} \approx 0.25$, consistent with triplet-like correlations on the rungs of the lower ladder.

However, in the spin-2 phase of the four-leg ladder with constant couplings of each kind,
the central rungs should also exhibit
$\langle C_{\text{rung}} \rangle_{l=2} \approx 0.25$, reflecting the ferromagnetic alignment of all four spins within each rung.
In Fig.~\ref{fig:coupledcorr}(b), we observe that $\langle C_{\text{rung}}\rangle_{l=2}$ changes smoothly from $\approx 0.25$ to negative values as $J_\times'$ is weakened, indicating a crossover from a single effective spin-2 chain, where all rungs behave as triplets, to a regime in which the triplet states on the two ladders couple antiferromagnetically, thereby suppressing triplet correlations on the central rungs.

In Fig.~\ref{fig:coupledcorr}(c), we show the correlations between spins
coupled by all types of $J$ interactions for $J_\perp'=0.5$ and
$J_\times=0.7$, with $J_\times' = 0.1,\, 0.29,$ and $0.5$, which correspond
approximately to $0.14J_\times$, $0.41J_\times$, and $0.71J_\times$,
respectively. In all three cases, the spins along the legs display
short-range antiferromagnetic order. For $J_\times' \approx 0.14J_\times$,
the two spin-1 sites within each rung are antiferromagnetically correlated,
since $J_\perp'$ is sufficiently larger than $J_\times'$. For
$J_\times' \approx 0.41J_\times$, the correlations between the central spins
satisfy $\langle C_{\text{rung}} \rangle_{l=2} \approx 0$, placing the system
near the crossover point; the spin-1 degrees of freedom of each ladder become
nearly uncorrelated, and the system behaves essentially as two decoupled
spin-1 chains, or even as a single effective chain, due to correlations that
develop between the edge spins of the two 2-leg ladders. In particular, triplet
correlations appear at the edge along legs $l=2$ and $l=3$. {A definitive characterization, however, of this effective spin-1
regime via its entanglement spectrum would require a careful
treatment of the edge geometry and the associated finite-size
effects, and is left for future work.} Finally, for
$J_\times' \approx 0.71J_\times$, an approximate spin-2-chain phase is
stabilized, characterized by strong triplet correlations between the central
spins on each rung in the bulk.

\begin{figure}
    \centering
    \includegraphics[width=0.4\textwidth]{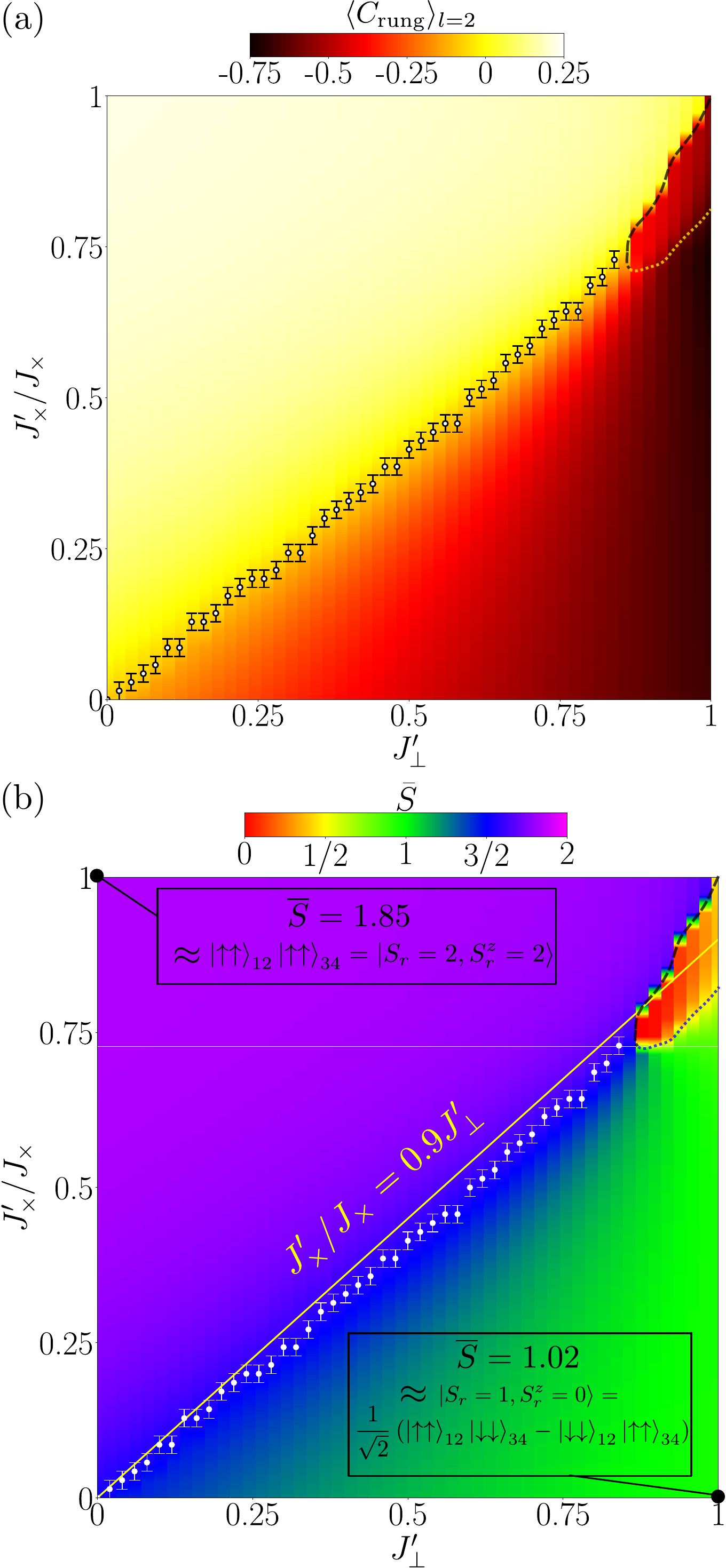}
    \caption{DMRG results for (a) the rung spin--spin correlations
$\langle C_{\text{rung}}\rangle_{l=2}$ and (b) the average effective total
rung spin $\overline{S}$ for $L=32$, $J_\perp=1$, $J_\parallel=0.8$, and
$J_\times=0.7$, shown as color maps in the $(J_\perp',\, J_\times'/J_\times)$
plane. In both panels, circles mark the points where
$\langle C_{\text{rung}}\rangle_{l=2}=0$, indicating the crossover line,
while dashed and dotted lines are guides to the first-order transition lines.
In panel (b), we also indicate the values of $\overline{S}$ and the approximate
rung state at the points $(0,1)$ and $(1,0)$, and display the straight line
$J_\times'/J_\times = 0.9\, J_\perp'$ as an estimate of the crossover line
(see text).
}
    \label{fig:diagramcoupled07}
\end{figure}

In Fig.~\ref{fig:diagramcoupled07}, we present the average central rung
correlation $\langle C_{\text{rung}}\rangle_{l=2}$ and the average rung
total spin $\overline{S}$ in the $(J'_\times/J_\times,\, J'_\perp)$ plane,
for $J_\perp=1$, $J_\parallel=0.8$, and $J_\times=0.7$. In both panels,
we indicate the crossover line, defined by
$\langle C_{\text{rung}}\rangle_{l=2}=0$, as well as the first-order
transition line. As shown in Fig.~\ref{fig:diagramcoupled07}, one finds
$\overline{S}=1.85$ at $(J'_\perp,J'_\times/J_\times)=(0,1)$, and
$\overline{S}=1.02$ at $(1,0)$. The value of $\overline{S}$ changes smoothly
between these two limits as the relative orientation of the triplets in the
two 2-leg ladders evolves with $J'_\times$ and $J'_\perp$. At the crossover
line, a coherent superposition of rung states with total spin 2 and total
spin 1, $\ket{C}=\frac{1}{\sqrt{2}}\left(\ket{S_r=2}+\ket{S_r=1}\right),$
yields $\bra{C}\mathbf{S}_r^2\ket{C}=4.0$, corresponding to an effective
rung spin $(S_r)_C = 1.56$, in good agreement with the DMRG results, which
give $1.47 < \overline{S} < 1.64$ along the crossover line.

\begin{figure}
    \centering
    \includegraphics[width=0.45\textwidth]{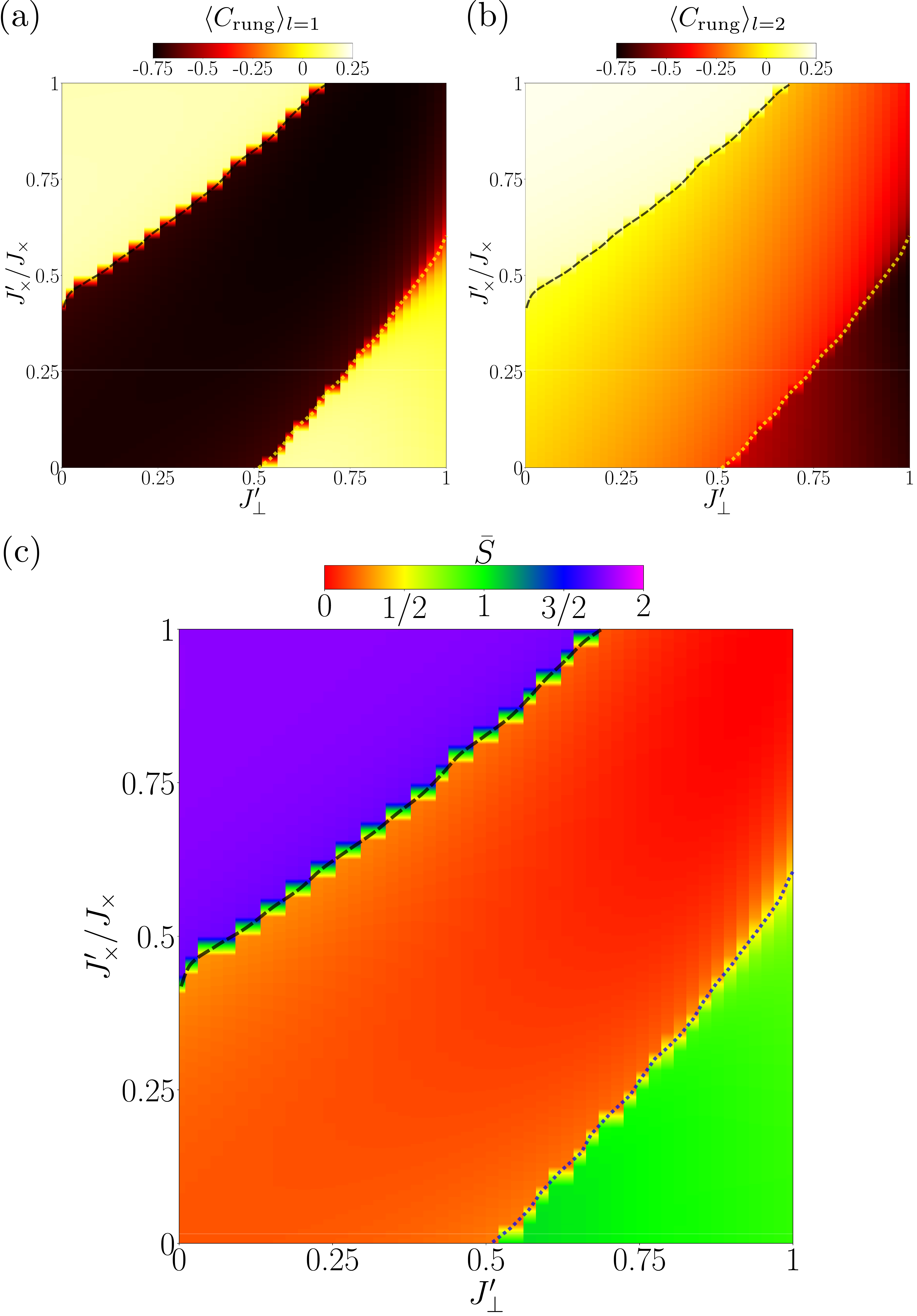}
    \caption{DMRG results for the average rung spin--spin correlations
$\langle C_{\text{rung}}\rangle_l$ and the average effective total rung
spin $\overline{S}$ for $L=32$, $J_\perp=1$, $J_\parallel=0.8$, and
$J_\times=0.64$. Panels show color maps of
(a) $\langle C_{\text{rung}}\rangle_{l=1}$,
(b) $\langle C_{\text{rung}}\rangle_{l=2}$, and
(c) $\overline{S}$ in the $(J_\perp',\, J_\times'/J_\times)$ plane.
Dashed and dotted lines serve as guides to the first-order transition lines.}
    \label{fig:diagramcoupled64}
\end{figure}

Although $\overline{S}$ does not remain fixed at $\approx 2$ above the
crossover line, a good estimate for this line can still be obtained by
projecting the Hamiltonian onto the subspace with $S_r=2$, and determining
the point at which its energy becomes equal to that of two decoupled 2-leg
ladders, where the $S_r=2$ state is degenerate with the $S_r=0$ and
$S_r=1$ states. This projection is performed using
Eq.~(\ref{eq:projection}), yielding

\begin{equation}
\mathcal{H}_{\{S_r=2\}} =
\frac{3}{4} J_\perp + \frac{3}{8} J_\perp' +
J_{\mathrm{eff}}'^{(\mathrm{spin\text{-}2})}
\sum_r \mathbf{T}_r \cdot \mathbf{T}_{r+1},
\label{eq:effspin2coupled}
\end{equation}

with

\begin{equation}
J_{\mathrm{eff}}'^{(\mathrm{spin\text{-}2})}
= \frac{1}{4}\left(J_\times + J_\parallel + \frac{J_\times'}{2}\right).
\end{equation}

The energy per { rung} derived from Eq.~(\ref{eq:effspin2coupled}),
$\varepsilon'_{J_\times,J_\perp}(J'_\times,J'_\perp)$, satisfies

\begin{equation}
\varepsilon'_{J_\times,J_\perp}(J'_\times,J'_\perp)
 - \varepsilon'_{J_\times,J_\perp}(0,0)
 = \frac{3}{8}J_\perp' + \frac{\tilde{\varepsilon} J_\times'}{8},
\end{equation}

where $\varepsilon'_{J_\times,J_\perp}(0,0)$ is the energy of two decoupled
ladders, and $\tilde{\varepsilon}=-4.76$ is the energy per site of a
spin-2 chain in units of $J_{\mathrm{eff}}'^{(\mathrm{spin\text{-}2})}$
\cite{USchollwock1995,schollwock_s_1996}.
The condition for effective decoupling,

\begin{equation}
\varepsilon'_{J_\times,J_\perp}(J'_\times,J'_\perp)
 - \varepsilon'_{J_\times,J_\perp}(0,0) = 0,
\end{equation}

leads to

\begin{equation}
\left(\frac{J'_\times}{J_\times}\right)_{\text{crossover}}
= -\frac{3 J'_\perp}{\tilde{\varepsilon} J_\times},
\label{eq:crossover}
\end{equation}

which, for the parameters used in Fig.~\ref{fig:diagramcoupled07}, reduces to
$\tfrac{J'_\times}{J_\times} \approx 0.9\, J'_\perp$. As shown in
Fig.~\ref{fig:diagramcoupled07}(b), this estimated crossover line agrees
remarkably well with the DMRG results, except in the region where the
spin-1 and spin-2 phases are separated by an intermediate regime with
reduced effective rung spin, $0 < \overline{S} < 1/2$. In particular, the
upper segment of this line (dashed) exhibits a strong first-order character,
as visible in Figs.~\ref{fig:coupledcorr}(a) and (b) for $J'_\perp=0.9$,
whereas the first-order transition weakens as $J'_\perp \to 1$ along the
lower portion (dotted) of the curve.

{ We note that the good agreement should be understood within the
semiclassical character of the projection
$\mathbf{S}_{r,l}=\tfrac{1}{4}\mathbf{T}_r$ used to derive
Eq.~(\ref{eq:effspin2coupled}): the same approximation is applied
on both sides of the comparison: the polarized rung in the
four-leg spin-2 phase and the polarized rung effectively realized
in the Haldane phase of each isolated two-leg ladder, so that
the estimate rests on a consistent semiclassical footing. The
crossover itself is a smooth feature, with intermediate effective
rung spin $1.47 \lesssim \overline{S} \lesssim 1.64$ rather than
$\overline{S}=2$, so a quantitative match is not guaranteed
\textit{a priori}; the close agreement with the DMRG data
nevertheless reflects this internal consistency of the
approximation.}

In Fig.~\ref{fig:diagramcoupled64}, we show the average rung correlations,
$\langle C_{\text{rung}}\rangle_{l=1}$ and $\langle C_{\text{rung}}\rangle_{l=2}$,
together with the average rung total spin $\overline{S}$ for $J_\perp=1$,
$J_\parallel=0.8$, and $J_\times=0.64$, in the $(J_\times'/J_\times,\, J_\perp')$
plane. For these couplings, an isolated frustrated ladder lies in a trivial
paramagnetic phase~\cite{almeida2023b}, while the corresponding 4-leg ladder
sits on the crossover line between the SR-AFM legs and SR-FM legs phases,
close to the first-order transition to the spin-2 phase, as shown in
Fig.~\ref{fig:diagram}(b).

In this case, strong singlet correlations on the rungs of each 2-leg ladder
persist throughout a broad region of the phase diagram
[Fig.~\ref{fig:diagramcoupled64}(a)]. Nonetheless, transitions to regimes
dominated by triplet correlations also occur. The quantity
$\langle C_{\text{rung}}\rangle_{l=2}$, shown in
Fig.~\ref{fig:diagramcoupled64}(b), decreases continuously from 0 to $-0.26$
along the diagonal $J_\times'/J_\times = J_\perp'$, and displays marked triplet
correlations in the other two phases as well.

A comparison between $\overline{S}$ for this case
[Fig.~\ref{fig:diagramcoupled64}(c)] and for $J_\times = 0.7$
[Fig.~\ref{fig:diagramcoupled07}(b)] reveals a clear suppression of the
crossover regime. The tiny region bounded by first-order transition lines in the
$J_\times = 0.7$ diagram expands into a substantially larger portion of the
phase diagram at $J_\times = 0.64$. However, similarly to the
$J_\times = 0.7$ case, the first-order transition becomes weak as
$J_\perp' \rightarrow 1$ along the dotted line.

\section{Summary and discussions}
\label{sec:summary}
In this work, we have investigated a four-leg spin-$1/2$ frustrated ladder
with diagonal couplings inside each plaquette, combining DMRG calculations
with effective-spin projections. In the uniform case, where all rungs and
diagonals are equivalent, we mapped out the ground-state phase diagram in the
$(J_\times,J_\parallel)$ plane and identified three main regimes: a phase with
short-range antiferromagnetic correlations along the legs (SR-AFM legs), a
phase with short-range ferromagnetic correlations (SR-FM legs), and a regime
where the ladder behaves as an effective spin-2 Heisenberg chain. The
crossover between the SR-AFM and SR-FM legs phases is signaled by a minimum of
the average effective total rung spin $\bar{S}$ and by dominant rung
correlations, whereas the spin-2 regime is separated from the SR-AFM and
SR-FM legs phases by a first-order transition line.

The spin-2 phase was characterized in more detail by analyzing the spin gap,
the string order parameter, and local magnetization profiles. The extrapolated
spin gap and nonvanishing string order parameter are consistent with the
expectations for a spin-2 Haldane chain, and the edge-localized excitations in
the $S^z=1$ sector further support this effective description. A simple
projection of the microscopic Hamiltonian onto the subspace where all four
spins in a rung form a spin-2 multiplet leads to an effective spin-2 chain
with coupling $J_{\mathrm{eff}}^{(\mathrm{spin\text{-}2})}$, whose excitation
gap $\Delta^{(\mathrm{eff})}$ agrees very well with the DMRG results in the
spin-2 region of the phase diagram. {We stress that the effective rung spin $\bar{S}$ is an emergent quantity that varies across the phase diagram, taking values from $0$ to $2$.}

In the second part of the paper, we reinterpreted the four-leg ladder as two
frustrated two-leg ladders coupled by intra-rung and diagonal interladder
interactions, $J'_\perp$ and $J'_\times$. This formulation allowed us to track
how the phases of an isolated two-leg ladder evolve into those of the four-leg
system. For $J_\parallel=0.8$, an isolated frustrated two-leg ladder realizes
either a trivial paramagnet ($J_\times \lesssim 0.7$) or an SPT spin-1 phase
($J_\times \gtrsim 0.7$). By turning on $J'_\perp$ and $J'_\times$, we showed
that these single-ladder states can be continuously deformed into the SR-AFM legs,
SR-FM legs, and spin-2 regimes of the uniform four-leg ladder.

For $J_\times=0.7$, where a single two-leg ladder is in the spin-1 SPT phase,
the coupled-ladder phase diagram in the $(J'_\times/J_\times,J'_\perp)$ plane
reveals a crossover line along which the central rung correlation
$\langle C_{\text{rung}}\rangle_{l=2}$ vanishes and the average effective rung
spin takes intermediate values $1.47 \lesssim \bar{S} \lesssim 1.64$. This
behavior is well captured by an effective description in terms of a coherent
superposition of spin-1 and spin-2 rung states, and the location of the
crossover line is accurately reproduced by projecting the Hamiltonian onto the
$S_r=2$ subspace and equating its energy to that of two decoupled two-leg
ladders. The same analysis also clarifies the nature of the first-order
transition line separating the spin-2 and spin-1–like regimes and its
weakening as $J'_\perp \to 1$.

For $J_\times=0.64$, where an isolated two-leg ladder is in a trivial
paramagnetic phase, the corresponding coupled-ladder phase diagram shows that
singlet correlations on the rungs of each two-leg ladder remain robust over a
broad region, while triplet-dominated regimes still emerge upon increasing
$J'_\perp$ and $J'_\times$. Comparing the $J_\times=0.7$ and $J_\times=0.64$
cases, we find that the crossover region between spin-1–like and spin-2–like
behavior is significantly reshaped: the tiny domain bounded by first-order
transition lines near the uniform point at $J_\times=0.7$ expands into a much
larger portion of the phase diagram at $J_\times=0.64$, where the first-order
character becomes considerably weaker.

{
We note that several aspects of the results are not directly
anticipated by the Ising limit or by a naive interpolation between
the isolated-ladder and uniform-ladder limits. In the uniform
four-leg case, the AFM–FM degeneracy line of the Ising limit is
replaced by an extended crossover region with $\bar{S} \approx 0$,
while the transition to the spin-2 regime retains a first-order
character. The spin-2 phase itself is characterized by a finite
string order parameter, spin-1 edge states, and a gap that matches
the spin-2 chain prediction $\Delta^{(\mathrm{eff})}$ within the
accuracy of our DMRG calculations, consistent with a Haldane-type
phase rather than a classical spin-2 chain. In the coupled-ladder
case, the evolution between the two limits proceeds through an
intermediate regime with $\bar{S} \approx 1.56$, whose location is
captured by the spin-2 projection within a consistent semiclassical approximation, and whose extent depends
noticeably on the intraladder couplings.
}

Altogether, our results demonstrate how frustrated four-leg ladders provide a
controlled setting to interpolate between effective spin-1 and spin-2 chain
physics by tuning interladder couplings. The combination of DMRG and
projection techniques used here could be extended to study the response of
these systems to magnetic fields, disorder, or time-dependent driving, as well
as to analyze larger coupled-ladder arrays that more closely approach
two-dimensional frustrated magnets and their possible topological phases.

{
\begin{acknowledgments}
We acknowledge the support from Coordenação de Aperfeiçoamento de Pessoal de Nível Superior (CAPES), Grant No. 1575/2024, Conselho Nacional de Desenvolvimento Científico e Tecnológico (CNPq), through the Universal Call (Grant No.  407819/2024-0), and Fundação de Amparo à Ciência e Tecnologia do Estado de Pernambuco (FACEPE).
\end{acknowledgments}
}

\appendix
{
\section{Instability line of the isolated-rung regime toward the spin-2 phase}
\label{app:transition}

\begin{figure}
    \centering
    \includegraphics[width=0.4\textwidth]{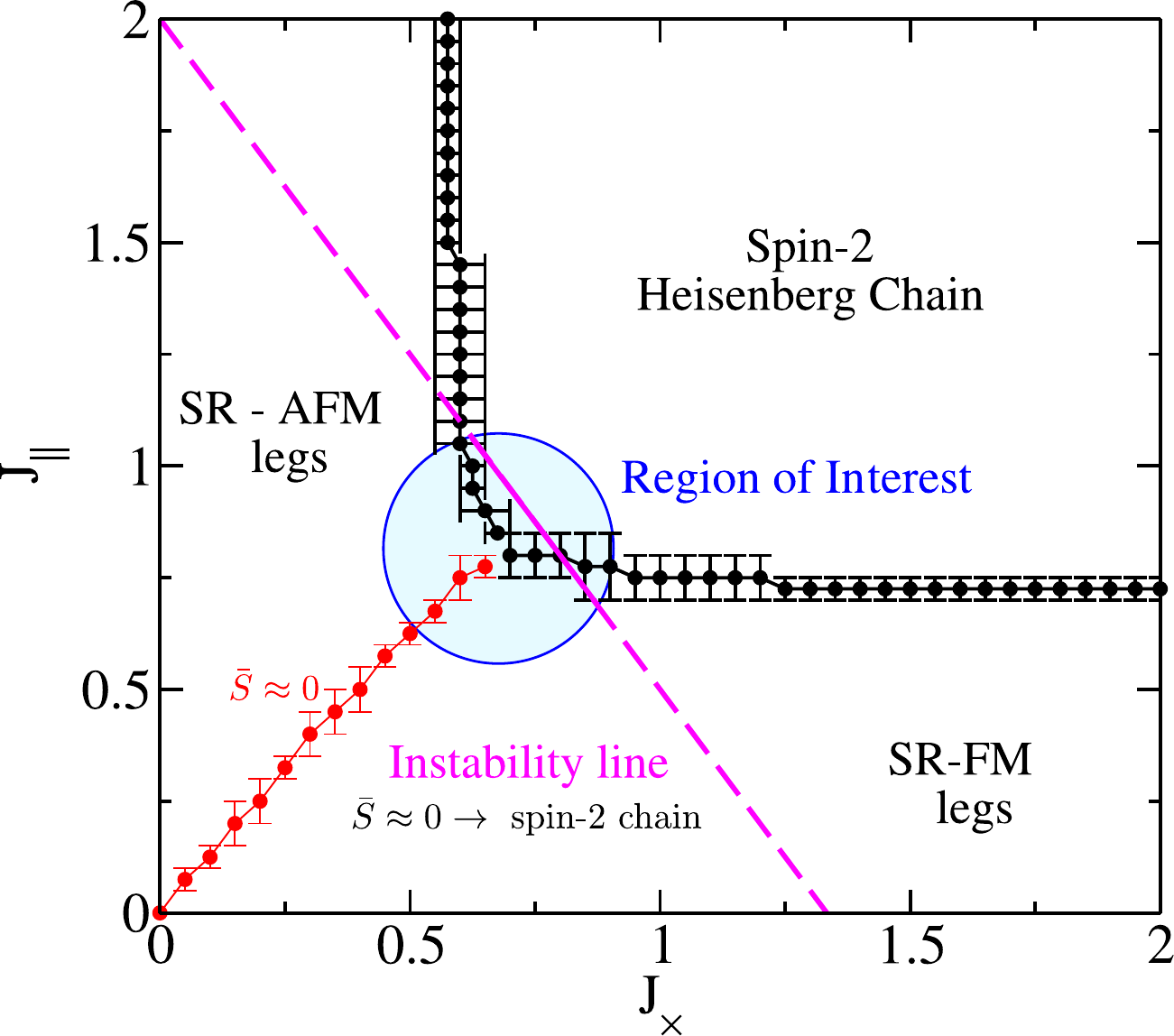}
    \caption{Phase boundaries of the uniform four-leg frustrated ladder
    (black and red symbols from Fig.~\ref{fig:diagram}) compared with the
    analytical instability line, Eq.~(\ref{eq:trans_estimate}) (dashed
    magenta line). The red symbols mark the crossover line ($\bar{S}\approx 0$).
    The shaded circle indicates the region of interest where the estimate
    is most reliable, near the junction of the crossover and first-order
    transition lines.}
    \label{fig:instability}
\end{figure}

Following Gelfand's approach to frustrated two-leg
ladders~\cite{Gelfand1991}, a simple estimate for the point at which
the first-order transition line meets the crossover line can be
obtained by comparing the energy per rung in the two competing
regimes. Along the crossover line, the rungs are nearly isolated and
the ground state is well approximated by a product of independent
rung ground states, with energy
$\varepsilon_0^{(S=0)} = -1.616\,J_\perp$ and
$\langle\mathbf{S}_r^2\rangle = 0$, obtained by exact diagonalization
of the four-spin rung Hamiltonian. In the spin-2 phase, the energy
per rung obtained from the projection in
Eq.~(\ref{eq:Hspin2eff}) is

\begin{equation}
\varepsilon_{\mathrm{spin\text{-}2}}
= \varepsilon_0^{(S=2)}
+ \tilde{\varepsilon}\,J_{\mathrm{eff}}^{(\mathrm{spin\text{-}2})},
\end{equation}

where $\tilde{\varepsilon}=-4.76$ is the ground-state energy per site
of the spin-2 Heisenberg chain~\cite{USchollwock1995,schollwock_s_1996}
and $J_{\mathrm{eff}}^{(\mathrm{spin\text{-}2})}$ is defined in
Eq.~(\ref{eq:Jeffspin2}). To place the two competing
regimes on equal footing, we evaluate the on-rung constant
$\varepsilon_0^{(S=2)}$ as the lowest energy of the four-spin rung
Hamiltonian in the $S_r=2$ sector: $\varepsilon_0^{(S=2)} = \tfrac{3}{4}\,J_\perp$. This places the
on-rung contribution on the same footing as the singlet side, where
$\varepsilon_0^{(S=0)}$ is also the exact lowest rung energy in its
total-spin sector. We note that the same constant evaluated through
the substitution $\mathbf{S}_{r,l}=\tfrac{1}{4}\mathbf{T}_r$ applied
directly to the on-rung bilinears yields
$\tfrac{3}{8}J_\perp$
per bond, i.e., a total $\tfrac{9}{8}J_\perp$ in the uniform limit, see Eq.(\ref{eq:effspin2coupled});
this overestimates the actual on-rung cost in the $S_r=2$ sector and
is therefore not the appropriate reference for the local-energy
comparison. Setting
$\varepsilon_{\mathrm{spin\text{-}2}} = \varepsilon_0^{(S=0)}$
gives the transition condition
\begin{equation}
J_{\mathrm{eff}}^{(\mathrm{spin\text{-}2})}
= \frac{1}{4}J_\parallel + \frac{3}{8}J_\times \approx 0.497,
\label{eq:trans_estimate}
\end{equation}
or equivalently $J_\parallel \approx 2 - 1.5\,J_\times$ (instability
line in Fig.~\ref{fig:instability}). Since this estimate neglects the
stabilization of the singlet-dominated phase by $J_\parallel$ and
$J_\times$, it provides a lower bound for the actual transition line.
As shown in Fig.~\ref{fig:instability}, the estimate approaches the
DMRG data near the junction of the crossover and first-order lines,
which is precisely the regime where the isolated-rung approximation
is most appropriate.

}
\bibliography{referencias}

\end{document}